\documentclass[twocolumn,,amsmath,amssymb]{revtex4}

\usepackage{graphicx}
\usepackage{dcolumn}
\usepackage{bm}
\newcommand{\etal}{\emph{et al.}}
\newcommand{\be}{\begin{equation}}
\newcommand{\ee}{\end{equation}}
\newcommand{\bfig}{\begin{figure}}
\newcommand{\efig}{\end{figure}}

\begin{document}      

\title{A gap-protected zero-Hall effect state in the quantum limit of the nonsymmorphic metal KHgSb
} 

\author{Sihang Liang$^1$}
\author{Satya Kushwaha$^2$}
\author{Tong Gao$^1$}
\author{Max Hirschberger$^1$}
\author{Jian Li$^1$}
\author{Zhijun Wang$^1$}
\author{Karoline Stolze$^2$}
\author{Brian Skinner$^3$}
\author{B. A. Bernevig$^1$}
\author{R. J. Cava$^2$}
\author{and N. P. Ong$^1$}
\affiliation{
{$^1$Department of Physics, $^2$Department of Chemistry, Princeton University, Princeton, NJ 08544\\
$^3$Department of Physics, Massachusetts Institute of Technology, Cambridge, Massachusetts 02139, USA}
}

\date{\today}      

\maketitle      
{\bf A recurring theme in topological matter is the protection of unusual electronic states by symmetry, for example, protection of the surface states in $Z_2$ topological insulators by time reversal symmetry~\cite{FuKaneMele,FuKane,QiHughesZhang}. Recently interest has turned to unusual surface states in the large class of nonsymmorphic materials~\cite{Parameswaran,Liu,Sato,YoungKane,FangFu,SatoGomi,Wang2016,Schoop}. In particular KHgSb is predicted to exhibit double quantum spin Hall (QSH) states~\cite{Wang2016}. Here we report observation of a novel feature of the Hall conductivity in KHgSb in strong magnetic field $\bf B$. In the quantum limit, the Hall conductivity is observed to fall exponentially to zero, but the diagonal conductivity is finite. A large gap protects this unusual zero-Hall state. We propose that, in this limit, the chemical potential drops into the bulk gap, intersecting equal numbers of right and left-moving QSH surface modes to produce the zero-Hall state.}

\begin{figure*}[t]
\includegraphics[width=18 cm]{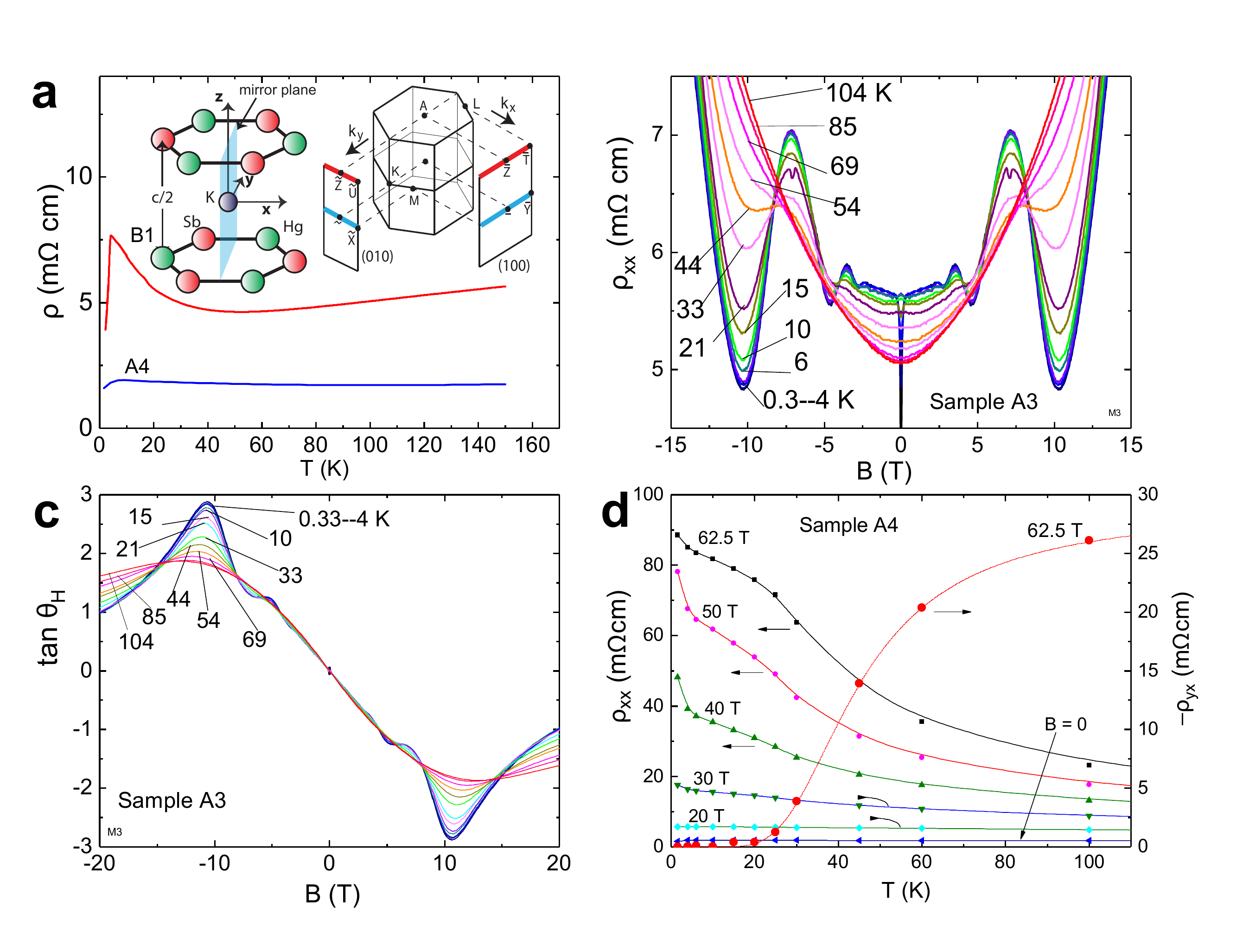}
\caption{\label{fig1} 
The quantum limit in KHgSb. Panel (a) shows the zero-$B$ resistivity $\rho$ vs. $T$ profile (Samples A4 and B1). The left inset is the crystal struture of KHgSb; symmetry points on the surface BZ are labelled in the right inset. Blue and red lines are mirror planes $\bar{M_z}$ at $k_z = 0, \pi/c$. Panel (b) shows in expanded scale the SdH oscillations for 0.3 $<T<$ 104 K (Sample A3, $\bf B\parallel\hat{z}$). Indexing the oscillations reveals that $\mu$ enters the LLL at the deep minimum near 10 T. Damping of the SdH amplitudes yields an in-plane mass $m_a$ = 0.05$m_e$. Panel (c) plots the field profiles of $\tan\theta_H$ at selected $T$ (in A3). At the peak ($\sim$11 T),  $\mu$ enters the LLL. Panel (d): In the extreme quantum limit ($B\gg B_{QL}$), the Hall resistivity $\rho_{yx}$ (red circles) and resistivity $\rho_{xx}$ (symbols) display divergent behaviors. As $T\to$2 K, $\rho_{xx}$ increases steeply implying a sharp decrease of the itinerant population. Paradoxically, the Hall resistivity $\rho_{yx}$ plunges to zero below 20 K (instead of diverging to very large values as expected from loss of itinerant carriers). These opposite trends imply a surface conduction channel in parallel with the bulk (see text). The saturation of $\rho_{xx}$ at 62 T as $T\to$2 K also constitutes evidence for the surface conduction mode. The solid red curve is a fit to $\rho_{yx}$ using Eq. \ref{sxy}; curves drawn through $\rho_{xx}$ symbols are guides to the eye. 
}
\end{figure*}

KHgSb crystallizes in the nonsymmorphic space group $D^4_{6h}(P6_3 /mmc)$. The Hg and Sb ions define honeycomb layers with $AB$ stacking (Fig. \ref{fig1}a, inset)~\cite{Wang2016}. The combination of strong spin-orbit coupling, inversion symmetry and band inversion leads to nontrivial topological properties (Supplementary Sec. S1). Because the mirror Chern number ${\cal C}_M$ = 2, we have double QSH states (2 left- and 2 right-moving modes) on each of the surfaces (100) and (010). The QSH states disperse along the blue lines in the inset in Fig. \ref{fig1}a, with velocity ${\bf v}_g\parallel$$\bf\hat{x}$ or $\bf\hat{y}$. At their intersections, left- and right-moving QSH states are protected against hybridization by mirror symmetry across the mirror plane $M_z$.

Crystals of KHgSb grow as plates with the broad faces normal to ${\bf\hat{z}}\parallel [001]$, and side faces identified with (100) and (010). Hall measurements in a field $\bf B\parallel \hat{z}$, with current in the $x$-$y$ plane can detect the QSH states, \emph{provided} the chemical potential $\mu$ lies inside the bulk gap. (Our measurements do not couple to the hourglass modes~\cite{HongDing,YLChen} because they disperse along $\tilde{\Gamma}\tilde{Z}$ with ${\bf v}_g\parallel{\bf\hat{z}}$.)

We report results from two batches of crystals (Supplementary Secs. S2 and S3, and Table 1). In batch A (nominally undoped), Hg vacancies lead to a carrier density ($n$-type) $n\sim 9.5\times 10^{17}$ cm$^{-3}$ (determined from the weak-$B$ Hall effect). In batch B, Bi dopants were added to reduce $n$ by a factor of 4. In both batches, $\mu$ lies low in the conduction band when $B=0$. In batch A, the in-plane resistivity $\rho_a$ ($B=0$) is nearly independent of temperature $T$, with a mobility $\mu_e\sim$3,500 cm$^2$/Vs limited by dominant impurity scattering. In batch B, $\rho_a$ increases by $30-40 \%$ between 40 and 4 K (Fig. \ref{fig1}a). (The sharp downturns at 4 K are caused by trace superconductivity from exuded Hg ions at the crystal surface; they do not affect the conclusions.) Batch A crystals display strong Shubnikov de Haas (SdH) oscillations. In Fig. \ref{fig1}b, SdH oscillations in the resistivity $\rho_{xx}$ (with $\bf B\parallel \hat{z}$) are plotted at selected $T$. At the field $B_{QL}$ at which $\rho_{xx}$ has a deep minimum, $\mu$ enters the lowest Landau level (LLL). From the damping of the SdH amplitudes versus $T$ we infer a small in-plane mass $m_{a} = 0.05 m_e$ ($m_e$ is the free electron mass). The Fermi surface is highly elongated along $\bf\hat{z}$, implying a weak interlayer coupling (see Sec. S6 and Fig. S10).

A hint of interesting behavior in the LLL was first detected in the field profiles of the Hall angle $\tan\theta_H$ (Fig. \ref{fig1}c). Whereas $\tan\theta_H$ is $B$-linear and $T$ independent below 5 T, it varies strongly with $B$ and $T$ once $\mu$ enters the LLL (the peak occurs at $B_{QL}$). Further hints emerged from the $T$ dependencies of the resistivity $\rho_{xx}$ and Hall resistivity $\rho_{yx}$ in intense $\bf B\parallel \hat{z}$ (Fig. \ref{fig1}d). The prominent increase in $\rho_{xx}$ in a 63-T field (60-fold at 2 K) suggests a dramatic loss of carriers, which should cause the Hall resistivity $\rho_{yx}$ to diverge. Paradoxically, we find instead that $\rho_{yx}$ measured with $B$ fixed at 62.5 T (red circles) plunges steeply to zero below $\sim$40 K.

\begin{figure*}[t]
\includegraphics[width=16 cm]{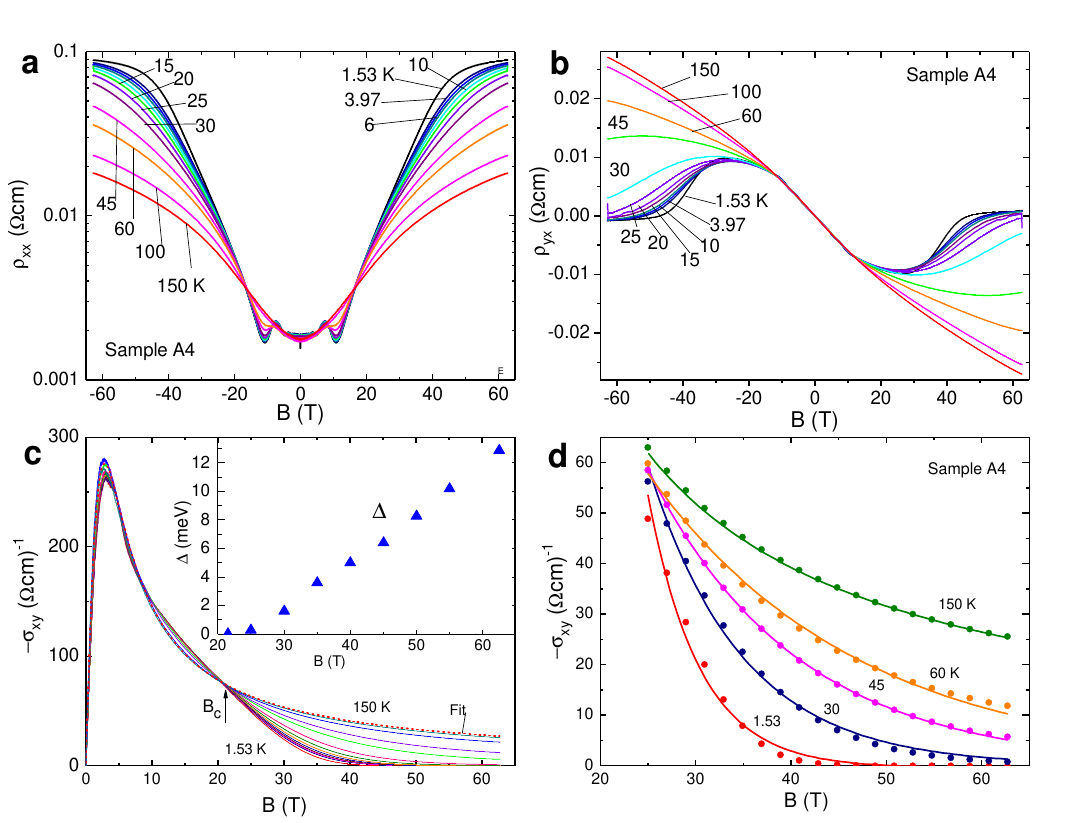}
\caption{\label{figA} 
Emergence of a zero-Hall state and gap in Sample A4. Panel (a) shows curves of $\rho_{xx}$ vs. $B$ for selected $T$ (semilog scale). At fields $B>\sim$11 T, $\rho_{xx}$ at 1.53 K increases steeply, but saturates above 60 T. Panel (b) displays the Hall resistivity curves $\rho_{yx}$ vs. $B$. Below 100 K, $\rho_{yx}$ displays strong $T$ dependence when $\mu$ enters the LLL. When $B$ exceeds $\sim$35 T, $\rho_{yx}$ falls steeply, consistent with thermal activation across a gap $\Delta(B)$. Panel (c) plots curves of $\sigma_{xy}$ vs. $B$ at selected $T$ (given in Panel (b)). The red dashed curve is the fit to Eq. \ref{sxy} with $n=n_0$ at 150 K. The strong $T$ dependence that appears above the onset field $B_c\sim$22 T (arrow) is well-described by Eq. \ref{sxy} with $n(T) = n_0\rm{e}^{-\Delta(B)/k_BT}$. The gap $\Delta(B)$ inferred from the fits increases to 13 meV at 62.5 T (inset). Panel (d) shows fits to Eq. \ref{sxy} (solid curves) for fields $B_c<B< 63$ T at selected $T$. At 150 K, we fix $n= n_0$ whereas at and below 60 K we use the thermally activated form for $n(T)$. 
}
\end{figure*}

To understand these features, we turn to the field profiles in Fig. \ref{figA} measured at low $T$ with $\bf B\parallel\hat{z}$. Using pulsed fields, we extended measurements of the resistivity $\rho_{xx}$ to 63 T, well beyond $B_{QL}\sim$11 T (Fig. \ref{figA}a). Whereas $\rho_{xx}$ is relatively flat below $B_{QL}$ (aside from the SdH oscillations), it shows a steep 60-fold increase above $B_{QL}$ (see curve at 1.53 K), consistent with a sharp decrease in the carrier density $n$, as noted above. However, above 45 T, $\rho_{xx}$ bends over to approach saturation instead of continuing to diverge.

Simultaneously, the Hall resistivity $\rho_{yx}$ displays a striking field profile (Fig. \ref{figA}b). At the lowest $T$ (1.53 K), $\rho_{yx}$ deviates sharply at $B_{QL}$ from the usual $B$-linear Hall dependence, describes a broad maximum and then plunges steeply to zero above 45 T, where it remains pinned up to the highest $B$= 63 T. As we raise $T$ from 1.5 to 10 K, the high-field curves of $\rho_{yx}$ strongly deviate from zero in a thermally activated way that defines an energy gap (Supplementary Figs. S1-S4).

In intense fields, it is best to analyze the Hall conductivity $\sigma_{xy}$. Inverting the matrix $\rho_{ij}$, we obtain the curves of $\sigma_{xy}(B)$ (Fig. \ref{figA}c). At 150 K, the measured $\sigma_{xy}$ closely follows the semiclassical form (shown as the dashed curve)
\be
\sigma_{xy} = n e\mu_e^2 B/[1+\mu_e^2B^2],
\label{sxy}
\ee
with $n$ the carrier density in the LLL and $e$ the charge. With $\mu_e$ = 3,500 cm$^2$/Vs, we have $\sigma_{xy}\to ne/B$ for $B>$20 T, i.e. $\sigma_{xy}$ depends only on $n$ because dissipation effects ($\mu_e$) cancel out. 

The key observation is that $\sigma_{xy}(T,B)$ becomes strongly $T$ dependent once $B$ exceeds an onset field $B_c\sim$22 T (arrow in Fig. \ref{figA}c). The strong $T$ dependence is well described by $n(T,B) = n_0{\rm{e}}^{-\Delta(B)/k_BT}$ with a $B$-dependent gap $\Delta(B)$ ($k_B$ is Boltzmann's constant). The close fits of $\sigma_{xy}$ are shown magnified in Fig. \ref{figA}d. (At each $B$, $\Delta(B)$ is uniquely obtained from the Arrhenius plot of $\sigma_{xy}$ with the prefactor $n_0 = 9.5\times 10^{17}$ cm$^{-3}$. The gap values are plotted in the inset in Fig. \ref{figA}c.). We stress that, above 20 T, $\sigma_{xy}$ depends only on $n(T,B)$ ($\mu_e$ cancels out). The activated behavior in $n(T,B)$ ensures that $\sigma_{xy}$ is pinned to zero in the limit $\Delta(B)/k_BT\gg 1$.

In our analysis, the conductivity $\sigma_{xx}$ initially increases with field just above $B_{QL}$ reflecting the increasing density of states (see Sec. S8). However, above $B_c$, $\sigma_{xx}$ falls steeply, and asymptotes to a $B$- and $T$-independent constant $\sigma^S$ which we identify with surface modes.  In the 7 batch A samples, $\sigma^S$ varies from 2.5 to 40 $(\Omega\mathrm{cm})^{-1}$ (or 0.03 and 0.1$e^2/h$ per HgSb layer). As emphasized above, the surface mode displays \emph{zero} Hall response. The finite $\sigma^S$ explains why $\rho_{yx}\to 0$ rather than $\infty$ in the limit $n\to 0$.

\begin{figure*}[t]
\includegraphics[width=16 cm]{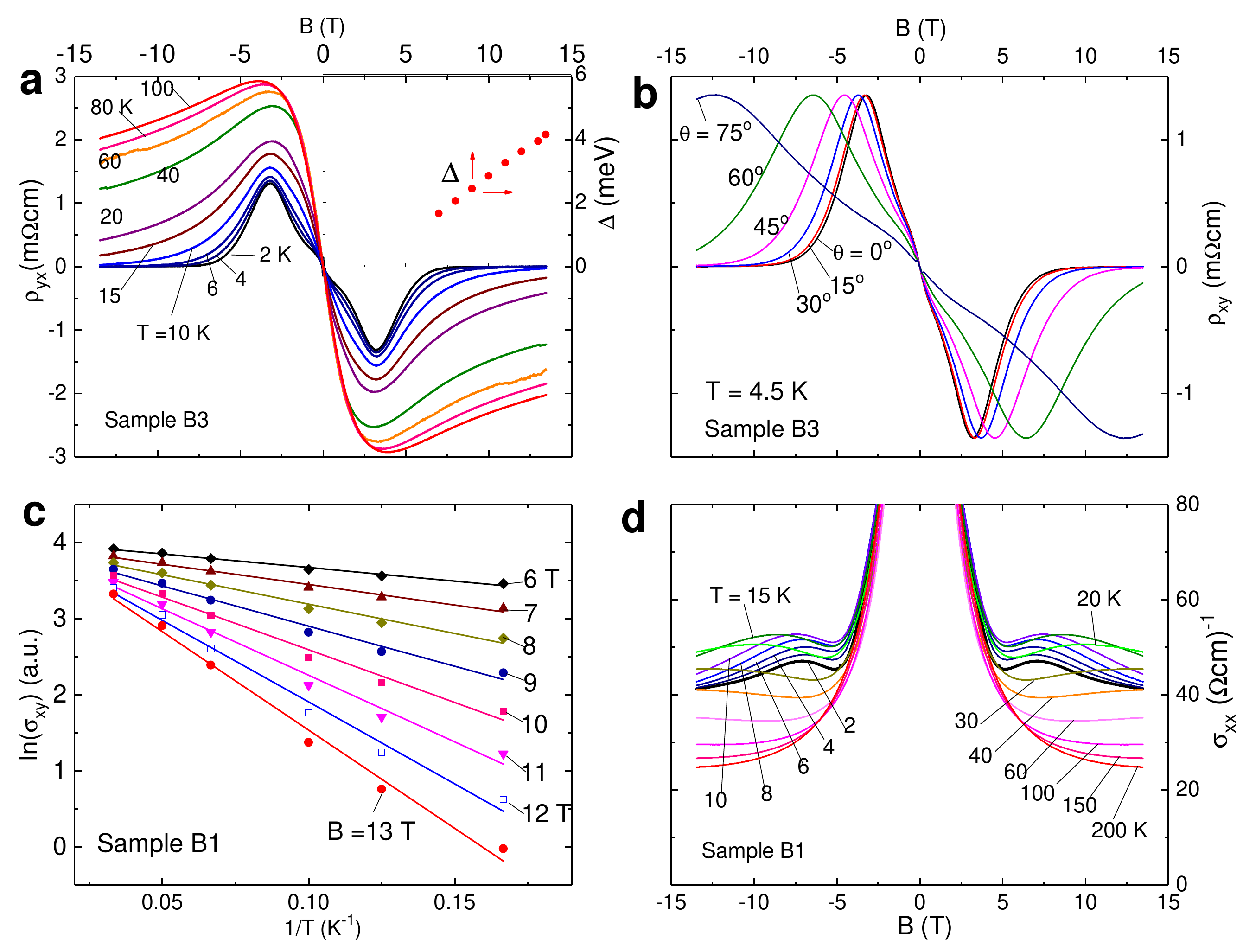}
\caption{\label{figB} 
The zero-Hall state in Bi-doped KHgSb (batch B). Panel (a) displays curves of the Hall resistivity $\rho_{yx}$ in B3. The field profiles are similar to Fig. \ref{figA}B except for the much lower field scales. The inferred gap $\Delta$ is plotted vs. $B$ in the inset. Panel (b) shows the effect of varying the tilt-angle $\theta$ of $\bf B$ relative to $\bf\hat{z}$ at 4.5 K. All the curves at $\theta<60^\circ$ collapse to a universal curve when plotted against $B_z$ (see Fig. S11). 
Panel (c) is a semilog plot of $\sigma_{xy}$ vs. $1/T$ at selected $B$ (in Sample B1). Curves of the conductivity $\sigma_{xx}$ vs. $B$ are displayed in Panel (d) for Sample B1. At 2 K (black curve), $\sigma_{xx}\to$ 41 $(\Omega \mathrm{cm})^{-1}$ or $\sim 0.1 e^2/h$ per mode above 10 T.
}
\end{figure*}

We also observe the zero-Hall state in Samples B1 and B3. Figure \ref{figB}a displays curves of $\rho_{yx}$ measured in Sample B3 (where $B_{QL}\sim$3 T, instead of 12 T). At 2 K, $\rho_{yx}$ approaches zero as an activated form when $B>$10 T. The gap $\Delta(B)$ inferred from the semilog plot of $\sigma_{yx}$ vs. $1/T$ (Fig. \ref{figB}c) is plotted in the inset in Panel (a). As shown in Fig. \ref{figB}c, the zero-Hall state survives to large field-tilt angles $\theta$ (see scaling plot in Supplementary Fig. S11). The similarity of the profiles of $\rho_{yx}$ in A4 (Fig. \ref{figA}b) and B3 (Fig. \ref{figB}b) implies that the zero-Hall state is intrinsic to the LLL (when $B>B_{QL}$). The large resistivity anisotropy ($\rho_{zz}/\rho_{xx}\sim$270; see Sec. S5 and Fig. S9), together with the highly elongated FS, imply a very weak interlayer hopping, consistent with dominant conduction via $\sigma^S$.

\begin{figure*}[t]
\includegraphics[width=16 cm]{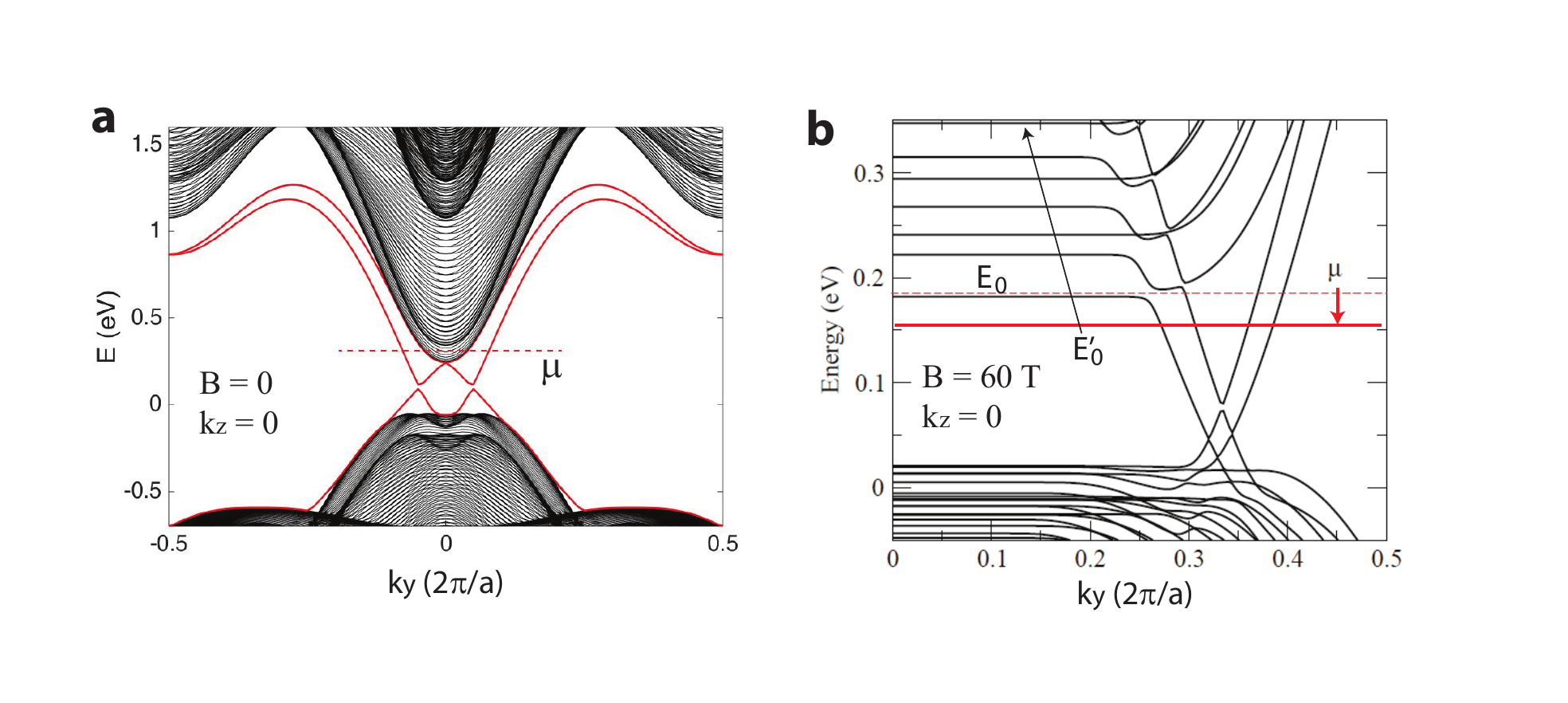}
\caption{\label{figbands} 
\emph{Ab initio} band structure of KHgSb near $\Gamma$ in $B=0$ and in strong $B$. In $B$ = 0 (Panel a), the bulk states (black curves) are separated by a direct gap of 320 meV. Pairs of right- and left-moving QSH states (red curves) traverse the gap. Panel (b) shows the LL spectrum in $B$ = 60 T with zig-zag termination on a (100) surface. In the conduction band, all LLs disperse upwards at the surface except for the lowest LL (at energy $E_0\sim$0.18 eV) and its partner produced by band folding at $k_z = \pi/c$ at energy $E'_0\sim$ 0.35 eV, which disperse downwards to form a pair of left-moving QSH states. Conversely, a pair of right-moving QSH states emerge from the valence band. In the quantum limit, magnetic freeze out lowers $\mu$ into the gap (dashed line to solid) where it intersects 2 left- and 2 right-moving QSH modes to give $\sigma_{xy} = 0$.
}
\end{figure*}

\emph{Ab initio} calculations reveal the special status of the LLL (Fig. \ref{figbands}, and Supplementary Sec. S7 and Fig. S12). In intense $B$ (60 T), the LLL in the conduction band and its partner level (at energies $E_0$=0.18 and $E_0'$ = 0.35 eV, respectively) disperse \emph{downwards} at the sample's edge, reflecting their hole-state origin (Fig. \ref{figbands}b). Together, these two edge states constitute the left-moving QSH modes. Conversely, a pair of right-moving QSH modes arise from the LL at the valence band maximum. Hence there exist 2 left- and 2 right-moving QSH modes. All 4 modes intersect $\mu$ if it lies in the bulk gap (solid line in Fig. \ref{figbands}b). By B$\ddot{\rm u}$ttiker-Landauer (BL) theory~\cite{Buttiker,Beenakker}, we must have $\sigma_{xy} = 0$, but $\sigma_{xx}$ will remain finite.

As noted, the disparate behaviors of $\sigma_{xy}$ and $\sigma_{xx}$ in the quantum limit suggest parallel conduction channels by bulk carriers and surface modes. In a conventional clean semimetal, $\mu$ is pinned at $E_0$ in the quantum limit (dashed line in Fig. \ref{figbands}b). However, if $\mu$ drops into the gap (i.e. itinerant bulk carriers vanish), the BL theory should apply.  One mechanism that produces such a downward shift in $\mu$ (as well as thermal activation in $n(T)$) is magnetic freezeout, long familiar in bulk semiconductors~\cite{Dyakonov,Drew,Mani} (Supplementary Sec. S8 and Fig. S13). Whereas the electrons are unbound in zero $B$, a strong $B$ deepens the binding energy to donors (Hg vacancies). At the field $B_c$, the binding energy becomes large enough to localize bulk carriers onto impurity states~\cite{Dyakonov}, so that $\mu$ falls below $E_0$ into the bulk gap (solid line in Fig.  \ref{figbands}b). When $\Delta/k_BT\gg 1$, the itinerant bulk population is completely suppressed. The entire current is then carried by the 4 QSH modes, so that the zero-Hall state with finite $\sigma_{xx}$ is realized.

We also consider whether a semiclassical 3-band Drude can explain our findings (details in Sec. S9 and Fig. S14). As discussed in Sec. S9, the key features here -- the sharp onset at $B_c$ of thermal activation across a gap $\Delta(B)$ and pinning of $\sigma_{xy}$ to zero up to 63 T -- lie well beyond the purview of semiclassical transport (see especially the plots of $\sigma_{xy}(B)$ in Figs. \ref{figA}c, \ref{figA}d and Fig. S15). Moreover, photoemission~\cite{HongDing,YLChen} and \emph{ab initio} band calculations do not see 3 bands. In addition, we distinguish our results from the anomalous Hall effect~\cite{Nagaosa} in ferromagnets and conventional semimetals (see Sec. S10).


\newpage

\vspace{1cm}\noindent

\vspace{5mm}\noindent
{\bf Acknowledgements} The research was supported by the Department of Energy (DE-SC0017863) and the Gordon and Betty Moore Foundation's EPiQS initiative through Grants GBMF4539 (to NPO) and GBMF-4412 (to RJC). The crystal growth and characterization were supported by the ARO MURI on topological insulators (Contract W911NF-12-1-0461), and by the U.S. National Science Foundation (Grant DMR 1420541). The high field experiments were performed at both the National High Magnetic Field Lab. NHMFL (Tallahassee) and at the Pulsed Field Facility (Los Alamos National Lab.). NHMFL is supported by the National Science Foundation Cooperative Agreement no. DMR-1157490, the State of Florida, and the US Department of Energy.  The theory work was supported by grants to BAB from the Department of Energy DE- SC0016239, Simons Investigator Award and NSF EAGER Award NOA - AWD1004957. B.S. is supported by the NSF STC "Center for Integrated Quantum Materials" under Cooperative Agreement No. DMR-1231319.

\vspace{3mm}
\noindent
{\bf Author contributions}\\
The experiment was planned by S.L., S.K., R.J.C., Z.W., B.A.B. and N.P.O. S.L. led the measurement effort with assistance from T.G. and M.H. The crystals were grown and characterized by S.K., K.S. and R.J.C. Theoretical support was provided by J.L., Z.W., B.A.B and B.S. The manuscript was written by N.P.O. with contributions from all authors.

\vspace{3mm}
\noindent
{\bf Additional Information}\\
Supplementary information is available in the online version of the paper.
Correspondence and requests for materials should be addressed to N.P.O.

\vspace{3mm}
\noindent
{\bf Competing financial interests}\\
The authors declare no competing financial interests.

\newpage

\end{document}